\documentclass[12pt,a4paper]{article}
\usepackage{cite}
\usepackage{url}
\usepackage{amsfonts}
\usepackage{amsmath}
\usepackage{amssymb}
\usepackage{graphicx}
\usepackage{epsfig}
\usepackage{lineno}
\usepackage[usenames]{color}
\usepackage{hyperref}
\usepackage{subfigure}
\usepackage[normalem]{ulem} 
\usepackage{color}
\usepackage{units}
\usepackage{xcolor}
\graphicspath{{./img/}}
\newif\ifhijing
\hijingtrue
\newif\iflatexdiff
\newif\ifcomment
\newcommand{\pp}           {pp}

\newcommand{\NN}           {NN}
\renewcommand{\AA}         {AA}

\newcommand{\pA}           {pA}
\newcommand{\dA}           {dA}
\newcommand{\pPb}          {pPb}
\newcommand{\AuAu}         {AuAu}
\newcommand{\PbPb}         {PbPb}
\newcommand{\XeXe}         {XeXe}
\newcommand{\OO}           {OO}

\newcommand{\Hpythia}      {HG-PYTHIA}
\newcommand{\Hijing}       {HIJING}
\newcommand{\Pythia}       {PYTHIA}
\newcommand{\MPI}          {MPI}
\newcommand{\pT}           {\ensuremath{p_{\rm T}}}
\newcommand{\pt}           {\pT}
\newcommand{\RAA}          {\ensuremath{R_{\rm AA}}}
\newcommand{\RCP}          {\ensuremath{R_{\rm CP}}}
\newcommand{\TAA}          {\ensuremath{T_{\rm AA}}}

\newcommand{\Ncoll}        {\ensuremath{N_{\rm coll}}}

\newcommand{\Nhard}        {\ensuremath{N_{\rm hard}}}
\newcommand{\Nhardnn}      {\ensuremath{N^{\rm hard}_{\rm NN}}}

\newcommand{\bnn}          {\ensuremath{b_{\rm NN}}}
\newcommand{\snn}          {\ensuremath{\sqrt{s_{\rm NN}}}}

\newcommand{\hrefurl}[1]   {\href{#1}{\url{#1}}}
\renewcommand{\Ref}[1]       {Ref.~\cite{#1}}

\newcommand{\Tab}[1]       {Tab.~\ref{#1}}
\newcommand{\Fig}[1]       {Fig.~\ref{#1}}
\newcommand{\Figs}[2]      {Figs.~\ref{#1} and \ref{#2}}
\newcommand{\Eq}[1]        {Eq.~\ref{#1}}

\newcommand{\co}[1]        {}

\begin{document}
\title{Absence of jet quenching\\in peripheral nucleus--nucleus collisions}
\author{
       Constantin Loizides \\
       \small{\textit{LBNL, Berkeley, CA, 94720, USA}} \\
       Andreas Morsch \\ 
       \small{\textit{CERN, 1211 Geneva 23, Switzerland}} \\
}
\maketitle
\begin{abstract}
Medium effects on the production of high-$\pT$ particles in nucleus--nucleus~(\AA) collisions are generally quantified by the nuclear modification factor~($\RAA$), defined to be unity in absence of nuclear effects.
Modeling particle production including a nucleon--nucleon impact parameter dependence, we demonstrate that $\RAA$ at midrapidity in peripheral \AA\ collisions can be significantly affected by event selection and geometry biases.
Even without jet quenching and shadowing, these biases cause an apparent suppression for $\RAA$ in peripheral collisions, and are relevant for all types of hard probes and all collision energies. 
Our studies indicate that calculations of jet quenching in peripheral \AA\ collisions should account for the biases, or else they will overestimate the relevance of parton energy loss.
Similarly, expectations of parton energy loss in light--heavy collision systems based on comparison with apparent suppression seen in peripheral $\RAA$ should be revised.
Our interpretation of the peripheral $\RAA$ data would unify observations for lighter collision systems or lower energies where significant values of elliptic flow are observed despite the absence of strong jet quenching.
\end{abstract}

Medium effects on the production of high-$\pT$ particles are in general quantified by the nuclear modification factor
\begin{equation}
\RAA = \frac{Y_{\rm AA}}{\Ncoll Y_{\rm pp}} = \frac{Y_{\rm AA}}{\TAA\ \sigma_{\rm pp}}
\end{equation}
defined as the ratio of the per-event yield $Y_{\rm AA}$ measured in nucleus--nucleus~(\AA) collisions to the yield of an equivalent incoherent superposition of \Ncoll\ binary \pp\ collisions. 
The number of binary collisions depends on the overlap between the two colliding nuclei quantified by the nuclear overlap \TAA. 
It is expected that in the absence of nuclear effects \RAA\ is unity. 
However, strictly speaking this holds only for centrality integrated measurements. 
In this case \Ncoll\ is given by $\Ncoll = A^{2} \sigma_{\rm pp} / \sigma_{\rm AA}$, where $\sigma_{\rm pp}$ and  $\sigma_{\rm AA}$ are, respectively, the \pp\ and \AA\ inelastic cross-sections.
As will be outlined in the following in more detail, centrality classification can lead to the selection of \AA\ event samples for which the properties of the binary \NN\ collisions deviate from unbiased \pp\ collisions. 
In this case \RAA\ can deviate from unity even in the absence of nuclear  effects. 
There are two main origins for selection biases. 
Firstly, the spatial distribution of nucleons bound in nuclei in the plane transverse to the beam direction differs from those of protons in a beam leading to a bias on the \NN\ impact parameter. 
Secondly, centrality selection is based on measurements related to bulk, soft particle production. 
The selection can bias the mean multiplicity of individual \NN\ collisions and in case of a correlations between soft and hard particle production the yield of hard processes in \AA\ collisions.  

In the optical Glauber model~\cite{Miller:2007ri}, the nuclear overlap is obtained from the nuclear density distributions and the impact parameter $b$ between two nuclei, which is the only parameter characterising a collision.
Instead, Monte Carlo~(MC) Glauber models~\cite{Miller:2007ri} take into account collision-by-collision fluctuations at fixed impact parameter by allowing the positions of the nucleons in the nuclei to vary.
The number of binary collisions is obtained by assuming that the nucleons move on straight trajectories and a collision is counted if the nucleon--nucleon (\NN) impact parameter $\bnn$ is below a certain threshold~(usually given by the inelastic \NN\ cross section). 
For each simulated \AA\ collision the MC Glauber determines \Ncoll, and for each of the \Ncoll\ nucleon--nucleon collisions the impact parameter $\bnn^{i}$ and the respective collision position $(x^i, y^i)$ in the transverse plane. 
In this way, such calculations provide important information about the energy density distribution including its event-by-event fluctuations in the initial state of \AA\ collisions, which can be used as input for hydrodynamic calculations.
However, for the evaluation of the nuclear modification factor the information about the individual \NN\ collisions is usually ignored. 
An impact parameter dependent \NN\ profile can also be enabled in the GLISSANDO model~\cite{Rybczynski:2013yba}, but is not (yet) available in the widely-used standard Glauber MC~\cite{Alver:2008aq,Loizides:2014vua}.

In variance to the standard MC Glauber approach, the \Hijing\ model~\cite{Wang:1991hta} takes into account the possibility of multiple hard scatterings (multiple parton interactions) in the same \NN\ collision. 
As in MC models for \pp\ collisions like for example PYTHIA~\cite{Sjostrand:2006za}, the mean number of hard scatterings per collison depends on the \NN\ impact parameter.
While the \NN\ collisions are still modeled as incoherent, the production rate of hard processes is not proportional to $\Ncoll$ but to 
\begin{equation}
\Nhard=\Ncoll \cdot \left.\Nhardnn\right|_{\rm C} / \left<\Nhardnn\right>\,, 
\label{eq:nhard}
\end{equation}
where $\left.\Nhardnn\right|_{\rm C}$ is the average number of hard scatterings in a \NN\ collision for a given centrality selection and $\left<\Nhardnn\right>$ is its unbiased average value. 
Similarly, \Ref{Jia:2009mq} describes an extension of the optical Glauber model, in which the nuclear overlap function is obtained from a convolution between the product of the thickness functions of the two nuclei and the nucleon--nucleon overlap function. 

These extensions have important consequences for the \AA\ impact parameter dependence of hard processes. 
With respect to standard Glauber \Ncoll\ scaling, the number of hard processes is suppressed in peripheral collisions due to a simple geometrical bias.
With increasing AA impact parameter the phase space for collisions increases $\propto b$ whereas the nuclear density decreases leading to an increased probability for more peripheral than average \NN\ collisions.

A further consequence arises if the yield of hard and soft processes are correlated via the common $b_{\rm NN}$ and centrality selection is based on soft particle production~(multiplicity or summed energy)~\cite{Abelev:2013qoq,Adam:2014qja}. 
In this case for a given centrality class, the \NN\ collisions can be biased towards higher or lower than average impact parameters.
The event selection bias is in particular important when fluctuations of the centrality estimator due to $\bnn$ are of similar size as the dynamic range of \Ncoll, as in \pA\ collisions.

In contrast, centrality measurements based on zero-degree energy should not introduce any selection bias, while the geometric bias could still play a role. 
In the so called hybrid method, described in \Ref{Adam:2014qja}, the \pPb\ centrality selection is based on zero-degree neutral energy in the Pb-going directions~(slow neutrons) and \Ncoll\ is determined from the measured charged particle multiplicity $M$ according to $\Ncoll = \langle \Ncoll \rangle \cdot M /  \langle M \rangle$, 
where $\langle \Ncoll \rangle$ and  $\langle M \rangle$ are, respectively, the centrality averaged number of collisions and multiplicty. 
In case soft and hard particle yields are affected in the same way, the selection bias would cancel in the nuclear modification factor.

\begin{figure}[t!] 
 \centering 
 \includegraphics[width=0.85\linewidth]{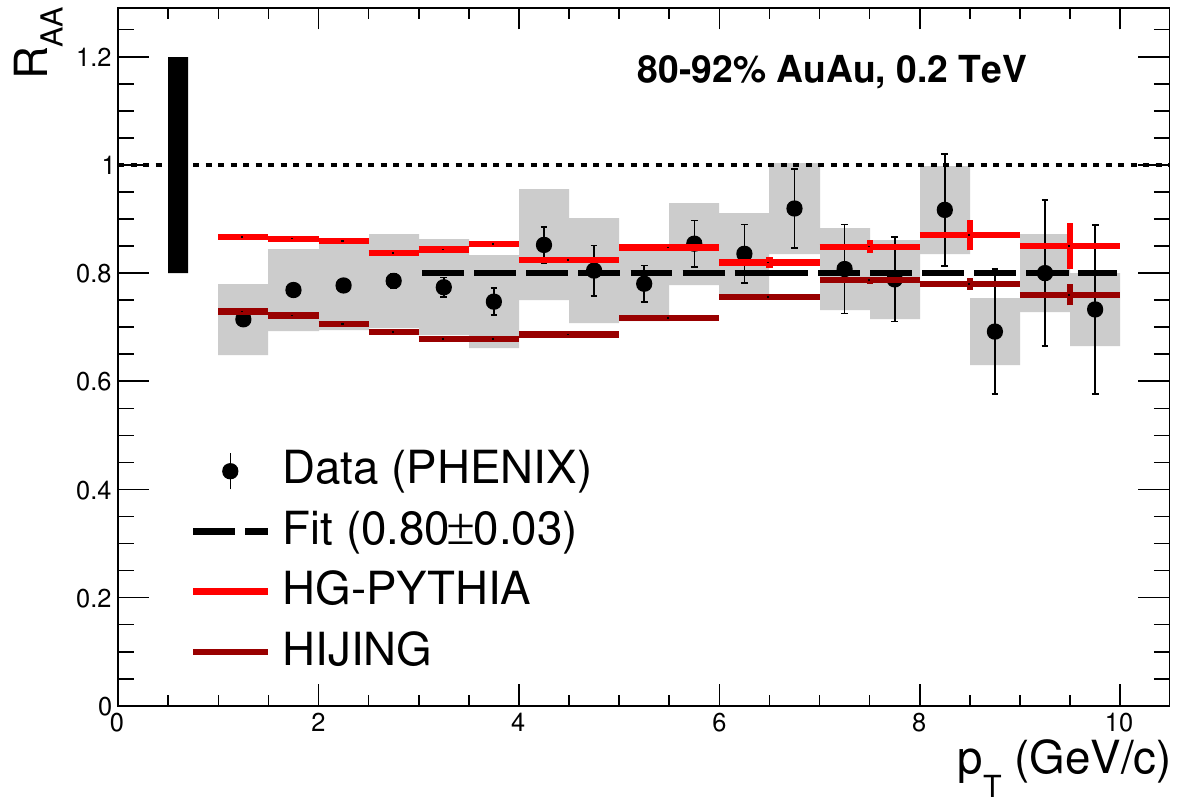}
 \caption{$\RAA$ versus $\pt$ in 80--92\% central \AuAu\ collisions at $\snn=0.2$~TeV. The PHENIX data from \cite{Adare:2008qa,Adare:2012wg}, which were averaged as explained in the text, are compared to \Hpythia\ and \Hijing\ calculations. For details, see text.} 
 \label{fig:rhic}
\end{figure}

\begin{figure}[h!] 
 \centering 
 \includegraphics[width=0.85\linewidth]{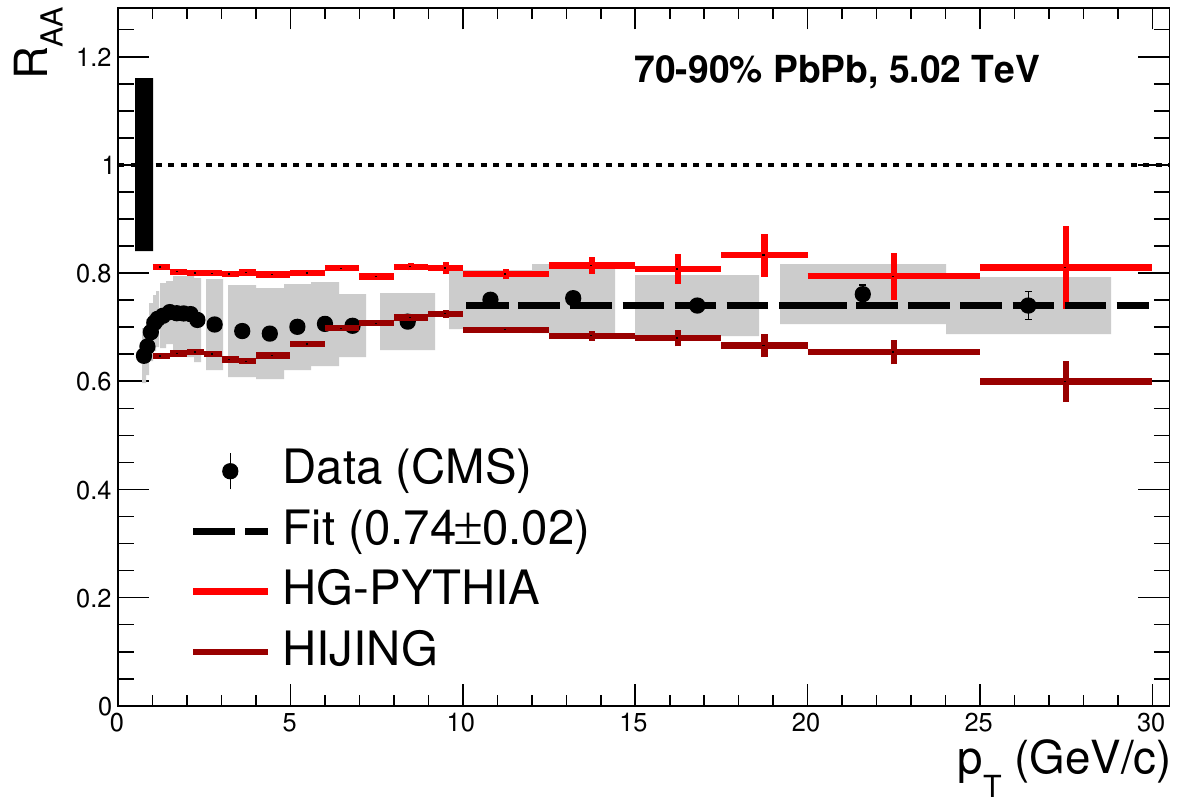}
 \caption{$\RAA$ versus $\pt$ in 70--90\% central \PbPb\ collisions at $\snn=5.02$~TeV. The CMS data from \cite{Khachatryan:2016odn} are compared to \Hpythia\ and \Hijing\ calculations. For details, see text.}
 \label{fig:lhc}
\end{figure}

In peripheral \AA\ collisions, one can expect the selection bias to be relevant, in addition to the geometric bias, as demonstrated in figure~8 of~\cite{Adam:2014qja}, which quantifies the ratio between the average multiplicity of the centrality estimator and the average multiplicity per average ancestor of the Glauber fit.
To illustrate its potential effect on peripheral \RAA\ we use PHENIX data in 80--92\% central \AuAu\ collisions at $\snn=0.2$~TeV~\cite{Adare:2008qa,Adare:2012wg} and CMS data in 70--90\% central \PbPb\ collisions at $\snn=5.02$~TeV~\cite{Khachatryan:2016odn}.
Above $5.25$~GeV/$c$ the PHENIX data from 2008 and 2012 were averaged using the quadratic sum of statistical and systematic uncertainties of the original measurements as weights.
The PHENIX data are shown in \Fig{fig:rhic} up to 10 GeV/$c$ and the CMS data in \Fig{fig:lhc} up to 30 GeV/$c$.
The error bars represent statistical, while the shaded boxes the systematic uncertainties. 
The vertical box around 1 at 0.5 GeV/$c$ denotes the global normalization uncertainty, which is dominated by the uncertainties on determining the centrality and $\Ncoll$~(or $\TAA$) from Glauber.
As indicated in the figures, constant functions were fit to the PHENIX and CMS data between $3$--$17$~GeV/$c$ and between $10$--$100$~GeV/$c$, respectively, yielding a value of $0.80\pm0.03$ and $0.74\pm0.02$ with a reduced $\chi^2<1$ using statistical and systematic uncertainties~(ignoring the normalization uncertainty) added in quadrature. 
Using a linear fit instead of a constant would in both cases result in a slope consistent with 0.
For \PbPb\ at $5.02$~TeV, this is distinctively different for the $50$--$70$\% centrality class, where $\RAA$ exhibits a significant slope of about $0.003$ GeV$^2$/$c^2$, indicating that $\RAA\sim1$ is reached at around $125$~GeV/$c$, although parton energy loss should play a stronger role than in the more peripheral class.

The data are compared to \Hijing~(v1.383) calculations without jet quenching and shadowing and a toy model called \Hpythia, which is based on the \Hijing\ Glauber model for the initial state and \Pythia~\cite{Sjostrand:2006za} as explained below.
Besides the jet quenching and shadowing settings\co{, which both were turned off}, all other settings in \Hijing\ were used as set by default, except the minimum $\pT$ of hard or semi-hard scatterings, which was set to $2.3$~(instead of $2.0$)~GeV for \PbPb\ collisions at $5.02$~TeV. 


\Hijing\ accounts for fluctuations of $\Nhardnn$ via an \NN\ overlap function $T_{\rm NN}$ that depends on $\bnn$. 
The probability for an inelastic \NN\ collision is given by
\begin{equation}
{\rm d} \sigma_{\rm inel} = 2\pi\, \bnn\, {\rm d}\bnn \, \left[1 - e^{-\left(\sigma_{\rm soft} + \sigma_{\rm hard}\right) \, T_{\rm NN} (\bnn)} \right]\,,
\end{equation}
where $\sigma_{\rm soft}$ is the geometrical soft cross-section of $57$~mb related to the nucleon size and $\sigma_{\rm hard}$ the energy-dependent pQCD cross-section for $2 \rightarrow 2$ parton scatterings.
At $\snn=0.2$ and $5.02$ TeV, $\sigma_{\rm hard}=11.7$ and $166.2$~mb, respectively, while below $0.04$~TeV it is only about $1$~mb. 
The \NN\ overlap is approximated by the Eikonal function as $T_{\rm NN} \propto \xi^3 \, K_{3}(\xi)$ with $\xi\propto\bnn/\sqrt{\sigma_{\rm soft}}$.
If there is at least one hard \NN\ collision, the number of hard scatterings or multiple parton interactions~(\MPI) per \NN\ interactions is distributed as
\begin{equation}
P(\Nhardnn) = \frac{{\langle \Nhardnn \rangle } ^{\Nhardnn}} {\Nhardnn!} e^{- \langle \Nhardnn \rangle }
\end{equation}
with the average number of hard scatterings determined by $\bnn$ as
\begin{equation}
 \langle \Nhardnn \rangle = \sigma_{\rm hard} \, T_{\rm NN} (\bnn) \,.
\end{equation}
For a soft \NN\ collision $\Nhardnn=0$.
The average number of hard collisions per \NN\ collisions decreases from $\left<\Nhardnn\right>=1.77$ at $\snn=5.02$~TeV to $0.28$ at $0.2$~TeV and becomes negligible~($<0.05$) below $0.04$~TeV.
The total number of hard scatterings for a \AA\ collision is then obtained by looping over all \NN\ collisions in the MC Glauber, i.e.\ $\Nhard=\sum_{i=1}^{\Ncoll} \left(\Nhardnn\right)_i$.

In \Hijing, $\RAA$ at high $\pT$ for central collisions is not unity even in absence of jet quenching and shadowing\co{~(and when selecting on $b$)}, possibly due to the implementation of adhoc energy--momentum conservation for large $\Nhardnn$ and dueto the treatment of multiple scattering to model the ``Cronin'' effect~\cite{Antreasyan:1978cw}. 
\ifhijing
This is illustrated in \Fig{fig:hijingcent}, which shows $\RAA$ versus $\pt$ in $0$--$5$\% central \AuAu\ and \PbPb\ collisions at $\snn=0.2$ and $5.02$~TeV, respectively, calculated using \Hijing\ without jet quenching and without shadowing.
At high $\pT$, a suppression by about a factor $2$ is obtained.
Such a strong suppression despite nuclear effects in the model calculation is not consistent with direct photon measurements~\cite{Afanasiev:2012dg,Chatrchyan:2012vq}, which find $\RAA\approx1$ with an uncertainty of about $\pm25$\% in central collisions. 
\fi

To overcome these features and to study a baseline corresponding to an incoherent and unconstrained superposition of \NN\ collisions, we developed \Hpythia.~\footnote{The code for HG-PYTHIA can be found at \url{https://github.com/abaty/HGPythia}.} 
For a given MC Glauber event calculated in \Hijing, we run \Pythia~(v6.28 with Perugia 2011 tune) to generate pp events with exactly $\left(\Nhardnn\right)_i$ 
multiple parton interactions for each \NN\ collision $i$, where $i$ runs from $0$ to \Ncoll . 
The generated particles from all \Pythia\ events are then combined and treated like a single \AA\ event in the further analysis.

The model calculations were performed at $\snn=0.2$~TeV for \AuAu\ and $5.02$~TeV for \PbPb\ collisions.
The respective pp references were obtained with the same \MPI\ model from \Hijing\ as for \AA\ collisions. 
The centrality selection was done similarly to the data, namely by ordering events based on the charged particle multiplicity in $3<|\eta|<4$ for the low and $2.5<|\eta|<5$ for the high energy data, respectively.
The multiplicity in the models approximately scales with $\Nhard$ and hence can not be expected to be in agreement with the data.
However, by ordering the events according to multiplicity for the centrality selection, this deficiency is not relevant for the present study.
The average number of binary collisions needed for the calculation of $\RAA$\co{ that can be compared to experimental data} was obtained directly from the model for the selected range in multiplicity and was found to be about $\Ncoll=4.1$ and $8.6$, respectively, similar to the values reported by the experiments. 

The results of the model calculations are compared to the data in \Figs{fig:rhic}{fig:lhc}.
As expected due to the selection bias, both calculations are below 1 even in absence of any initial or final state effect in the model.
The models describe the data at both collision energies, despite a factor of 25 difference in their center-of-mass energies. 
\Hpythia\ has the tendency to overshoot the data, while \Hijing\ to undershoot the data, and appears to decrease with increasing $\pT$.
Presumably this originates from the adhoc energy--momentum conservation and strong multiple scattering implemented in \Hijing, which is purposely neglected in \Hpythia, and hence the difference between the two models may be taken as a extreme range of model predictions.
The experimental results have large global normalization uncertainties of about $15$ to $20$\%, which however are largely correlated with those of the calculations, since in all cases the same types of MC Glauber simulation are used.
However, the PHENIX data may be affected by finite trigger inefficiency in the peripheral bin, whose correction as in the case of \dA\ collisions~\cite{Adare:2013nff} would work in the opposite direction than the multiplicity bias~\cite{Jamiepriv2}.

\begin{figure}[t] 
 \centering 
 \includegraphics[width=0.85\linewidth]{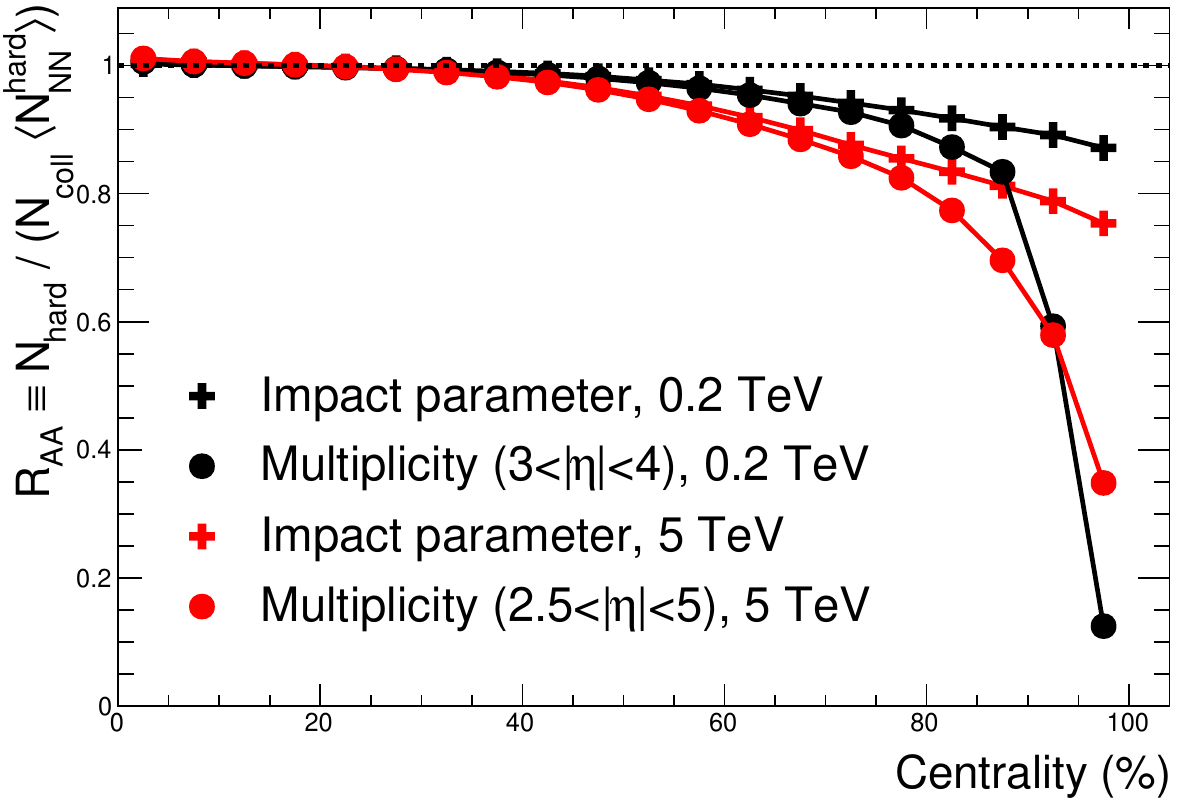}
 \caption{$\RAA$ versus centrality for \AuAu\ and \PbPb\ collisions at $\snn=0.2$ and $\snn=5.02$~TeV, respectively, calculated with \Hpythia, where $\RAA\equiv\Nhard/(\Ncoll\,\left<\Nhardnn\right>)$ by construction. For details, see text.}
 \label{fig:cent}
\end{figure}

Finally, we discuss the relative importance of the geometry and selection biases on $\RAA$. 
In \Hpythia, $\RAA$ at high $\pT$ approaches $\left.\Nhardnn\right|_{\rm C}=\Nhard/\Ncoll/\left<\Nhardnn\right>$ from \Eq{eq:nhard} by construction. 
The ratio of $\Nhard/\Ncoll$ for \AuAu\ and \PbPb\ collisions divided by the average number of $\Nhard$ in \pp\ collisions at $\snn=0.2$ and $\snn=5.02$~TeV, respectively, was calculated, and plotted versus centrality class in $5$\% intervals in \Fig{fig:cent}.
The events were either ordered according to impact parameter probing only the geometry bias, or according to charged particle multiplicity in the same rapidity ranges as before, probing both the geometry and the selection biases.
It turns out that the geometry bias sets in at mid-central collisions, reaching up to $10$ to $20$\% for most peripheral collisions.
The additional effect of the selection bias becomes noticeable above the 60\% percentile and is significant above the 80\% percentile, reaching up to $75\%$ and $40$\% for most peripheral collisions.
The onset of the selection bias is steeper for the lower collision energy, as expected if the fluctuations of the multiplicity and the dynamic range of $\Ncoll$ are similar.
For peripheral classes of $85$\% and beyond, another bias becomes increasingly important, namely the multiplicity selection in \AA\ collisions effectively cuts into the \pp\ cross section, suppressing the rate of hard processes relative to minimum bias \pp\ collisions.
While at center-of-mass energies of below 40~GeV~\cite{Luo:2015doi,Adamczyk:2017nof}, the effect from multiple interactions can be neglected, the latter, so called ``jet-veto''~\cite{Adam:2014qja}, bias is strongly enhanced due to auto-correlations, since centrality estimators at mid-rapidity were used.
Hence, the ratio of yields in central to peripheral collisions, $\RCP$, which is intended to quantify nuclear modification in absence of \pp\ reference data, would be larger than unity in absence of nuclear effects.

In conclusion, our model studies which employ particle production including a nucleon--nucleon impact parameter dependence in \AA\ collisions but no nuclear effects, demonstrate that $\RAA$ at midrapidity in peripheral \AA\ collisions can be significantly affected by event selection and geometry biases, similar to those previously discussed in the context of \pPb\ collisions at the LHC~\cite{Adam:2014qja}.
The apparent suppression induced by the biases for peripheral $\RAA$ results are present at all high energy collisions, and for very peripheral collisions expected to be stronger for lower energy collisions.
In principle, all hard probes should be similarly affected; examples are the most peripheral $\RAA$ of inclusive J/$\psi$~\cite{Adam:2016rdg} or jets~\cite{Khachatryan:2016jfl} at LHC energies. 
Furthermore, the discussion of the onset of parton energy loss studied with the RHIC beam energy scan~\cite{Luo:2015doi,Adamczyk:2017nof} should take the biases into account when considering the ratio of yields in central to peripheral collisions, which may lead to an artificial enhancement of $\RCP$ in absence of nuclear effects.
Calculations that attempt to address parton energy loss in peripheral collisions have to account for the effect, or else they will overestimate the suppression caused by parton energy loss.
In particular, to assess the multiplicity bias, the correlation between soft and hard particle production has to be modelled and centrality selection has to be performed in the same way as in the experiment.
For example, in the model calculation~\cite{Dainese:2004te}, the extracted transport coefficient changes by an order of magnitude from $0.1$~ GeV$^2/$fm, ie.\ similar to cold nuclear matter, for $\RAA\sim1$ in $80$--$92$\% to $1$~GeV$^2$/fm, ie. hot nuclear matter, for $\RAA\sim0.8$ in $60$--$70$\% \AuAu\ collisions at $\snn=0.2$~TeV.
Hence, our studies indicate that expectations of parton energy loss in light--heavy collision systems~\cite{Zhang:2013oca,Tywoniuk:2014hta,Shen:2016egw}, which were perhaps guided by comparing with apparent suppression seen in peripheral $\RAA$, should be revisited.

Our interpretation of the peripheral \RAA\ data also implies that large values of elliptic flow~($v_2$), as measured in peripheral \AA\ collisions~\cite{Chatrchyan:2013nka}, can arise in a system that does not exhibit jet quenching, and hence in accordance with similar observations in low-energy heavy-ion collisions~\cite{Adamczyk:2016gfs}, or in light--heavy or even light--light collisions~(see references in \cite{Loizides:2016tew}).

To further study the effect of the biases, it would be useful measure peripheral $\RAA$~(or integrated values above $\sim10$~GeV/$c$) in finer~($5$\% wide) centrality intervals. 
The trend of the progressively more peripheral $\RAA$ values should clearly demonstrate the effect of the biases, as one would expect the $\RAA$ values to decrease, and not to rise, as expected if parton energy loss decreases.
Further experimental studies, such as measurements of \RAA\ for reference processes like photon and intermediate vector boson for which no medium modification is expected, in peripheral \AA\ collisions, would be needed to experimentally constrain the biases, but they are not straightforward due to the reduced parton luminosity, and due to significant contamination from ultra-peripheral collisions and electro-magnetic processes.
Finally, semi-inclusive measurements as \cite{Adam:2015doa,Adamczyk:2017yhe} can be used to constrain the energy loss in peripheral \AA\ and light--heavy collision systems, as they do not rely on the modeling of the Glauber parameters.

\section*{Acknowledgments}
We thank Jamie~Nagle and Yen-Jie~Lee for fruitful discussions.
We thank Austin Baty for making available his code of \Hpythia~(which he derived from our ROOT-macro-based code end of 2019).
The work of C.\ Loizides is supported by the U.S. Department of Energy, Office of Science, Office of Nuclear Physics, under contract number DE-AC02-05CH11231.

\ifhijing
\begin{figure}[th!] 
 \centering 
 \includegraphics[width=0.85\linewidth]{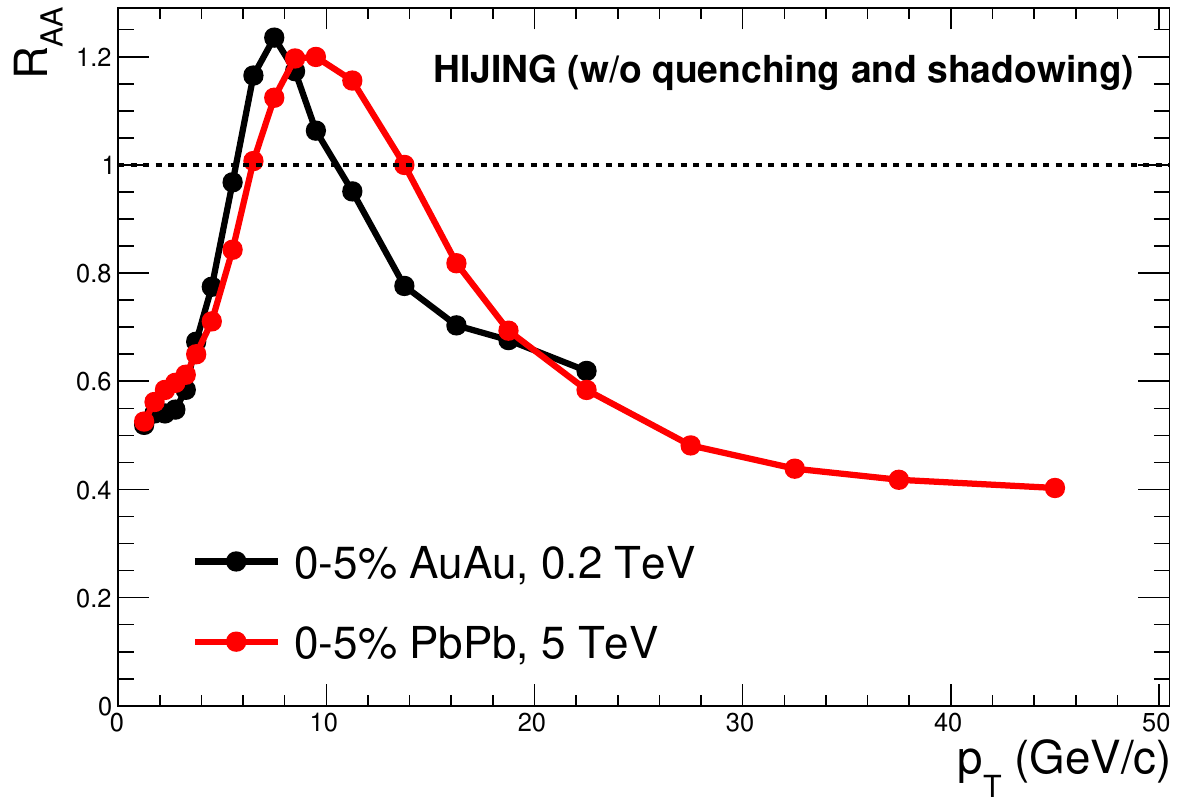}
 \caption{$\RAA$ versus $\pt$ in $0$--$5$\% central \AuAu\ and \PbPb\ collisions at $\snn=0.2$ and $5.02$~TeV, respectively calculated using \Hijing\ without jet quenching and without shadowing.}
 \label{fig:hijingcent}
\end{figure}
\fi

\bibliographystyle{utphys}
\bibliography{./biblio}

\providecommand{\href}[2]{#2}\begingroup\raggedright\begin{thebibliography}{10}

\bibitem{Miller:2007ri}
M.~L. Miller, K.~Reygers, S.~J. Sanders, and P.~Steinberg, ``{Glauber modeling
  in high energy nuclear collisions},''
  \href{http://dx.doi.org/10.1146/annurev.nucl.57.090506.123020}{{\em Ann. Rev.
  Nucl. Part. Sci.} {\bfseries 57} (2007) 205--243},
\href{http://arxiv.org/abs/nucl-ex/0701025}{{\ttfamily arXiv:nucl-ex/0701025
  [nucl-ex]}}.

\bibitem{Rybczynski:2013yba}
M.~Rybczynski, G.~Stefanek, W.~Broniowski, and P.~Bozek, ``{GLISSANDO 2 :
  GLauber Initial-State Simulation AND mOre…, ver. 2},''
  \href{http://dx.doi.org/10.1016/j.cpc.2014.02.016}{{\em Comput. Phys.
  Commun.} {\bfseries 185} (2014) 1759--1772},
\href{http://arxiv.org/abs/1310.5475}{{\ttfamily arXiv:1310.5475 [nucl-th]}}.

\bibitem{Alver:2008aq}
B.~Alver, M.~Baker, C.~Loizides, and P.~Steinberg, ``{The PHOBOS Glauber Monte
  Carlo},''
\href{http://arxiv.org/abs/0805.4411}{{\ttfamily arXiv:0805.4411 [nucl-ex]}}.

\bibitem{Loizides:2014vua}
C.~Loizides, J.~Nagle, and P.~Steinberg, ``{Improved version of the PHOBOS
  Glauber Monte Carlo},''
  \href{http://dx.doi.org/10.1016/j.softx.2015.05.001}{{\em SoftwareX}
  {\bfseries 1-2} (2015) 13--18},
\href{http://arxiv.org/abs/1408.2549}{{\ttfamily arXiv:1408.2549 [nucl-ex]}}.

\bibitem{Wang:1991hta}
X.-N. Wang and M.~Gyulassy, ``{HIJING: A Monte Carlo model for multiple jet
  production in \pp, \pA\ and \AA\ collisions},''
\href{http://dx.doi.org/10.1103/PhysRevD.44.3501}{{\em Phys. Rev.} {\bfseries
  D44} (1991) 3501--3516}.

\bibitem{Sjostrand:2006za}
T.~Sjostrand, S.~Mrenna, and P.~Z. Skands, ``{PYTHIA 6.4 Physics and Manual},''
  \href{http://dx.doi.org/10.1088/1126-6708/2006/05/026}{{\em JHEP} {\bfseries
  05} (2006) 026},
\href{http://arxiv.org/abs/hep-ph/0603175}{{\ttfamily arXiv:hep-ph/0603175
  [hep-ph]}}.

\bibitem{Jia:2009mq}
J.~Jia, ``{Influence of the nucleon-nucleon collision geometry on the
  determination of the nuclear modification factor for nucleon-nucleus and
  nucleus-nucleus collisions},''
  \href{http://dx.doi.org/10.1016/j.physletb.2009.10.044}{{\em Phys. Lett.}
  {\bfseries B681} (2009) 320--325},
\href{http://arxiv.org/abs/0907.4175}{{\ttfamily arXiv:0907.4175 [nucl-th]}}.

\bibitem{Abelev:2013qoq}
{\bfseries ALICE} Collaboration, B.~Abelev {\em et~al.}, ``{Centrality
  determination of \PbPb\ collisions at $\snn = 2.76$ TeV with ALICE},''
  \href{http://dx.doi.org/10.1103/PhysRevC.88.044909}{{\em Phys. Rev.}
  {\bfseries C88} no.~4, (2013) 044909},
\href{http://arxiv.org/abs/1301.4361}{{\ttfamily arXiv:1301.4361 [nucl-ex]}}.

\bibitem{Adam:2014qja}
{\bfseries ALICE} Collaboration, J.~Adam {\em et~al.}, ``{Centrality dependence
  of particle production in \pPb\ collisions at $\snn = 5.02$ TeV},''
  \href{http://dx.doi.org/10.1103/PhysRevC.91.064905}{{\em Phys. Rev.}
  {\bfseries C91} no.~6, (2015) 064905},
\href{http://arxiv.org/abs/1412.6828}{{\ttfamily arXiv:1412.6828 [nucl-ex]}}.

\bibitem{Adare:2008qa}
{\bfseries PHENIX} Collaboration, A.~Adare {\em et~al.}, ``{Suppression pattern
  of neutral pions at high transverse momentum in \AuAu\ collisions at
  $\snn=200$ GeV and constraints on medium transport coefficients},''
  \href{http://dx.doi.org/10.1103/PhysRevLett.101.232301}{{\em Phys. Rev.
  Lett.} {\bfseries 101} (2008) 232301},
\href{http://arxiv.org/abs/0801.4020}{{\ttfamily arXiv:0801.4020 [nucl-ex]}}.

\bibitem{Adare:2012wg}
{\bfseries PHENIX} Collaboration, A.~Adare {\em et~al.}, ``{Neutral pion
  production with respect to centrality and reaction plane in \AuAu\ collisions
  at $\snn = 200$ GeV},''
  \href{http://dx.doi.org/10.1103/PhysRevC.87.034911}{{\em Phys. Rev.}
  {\bfseries C87} no.~3, (2013) 034911},
\href{http://arxiv.org/abs/1208.2254}{{\ttfamily arXiv:1208.2254 [nucl-ex]}}.

\bibitem{Khachatryan:2016odn}
{\bfseries CMS} Collaboration, V.~Khachatryan {\em et~al.}, ``{Charged-particle
  nuclear modification factors in \PbPb\ and \pPb\ collisions at $\snn = 5.02$
  TeV},'' \href{http://dx.doi.org/10.1007/JHEP04(2017)039}{{\em JHEP}
  {\bfseries 04} (2017) 039},
\href{http://arxiv.org/abs/1611.01664}{{\ttfamily arXiv:1611.01664 [nucl-ex]}}.

\bibitem{Antreasyan:1978cw}
D.~Antreasyan, J.~W. Cronin, H.~J. Frisch, M.~J. Shochet, L.~Kluberg, P.~A.
  Piroue, and R.~L. Sumner, ``{Production of hadrons at large transverse
  momentum in 200, 300 and 400 GeV \pp\ and \pn\ collisions},''
\href{http://dx.doi.org/10.1103/PhysRevD.19.764}{{\em Phys. Rev.} {\bfseries
  D19} (1979) 764--778}.

\bibitem{Afanasiev:2012dg}
{\bfseries PHENIX} Collaboration, S.~Afanasiev {\em et~al.}, ``{Measurement of
  direct photons in \AuAu\ collisions at $\snn=200$ GeV},''
  \href{http://dx.doi.org/10.1103/PhysRevLett.109.152302}{{\em Phys. Rev.
  Lett.} {\bfseries 109} (2012) 152302},
\href{http://arxiv.org/abs/1205.5759}{{\ttfamily arXiv:1205.5759 [nucl-ex]}}.

\bibitem{Chatrchyan:2012vq}
{\bfseries CMS} Collaboration, S.~Chatrchyan {\em et~al.}, ``{Measurement of
  isolated photon production in pp and PbPb collisions at $\snn=2.76$ TeV},''
  \href{http://dx.doi.org/10.1016/j.physletb.2012.02.077}{{\em Phys. Lett.}
  {\bfseries B710} (2012) 256--277},
\href{http://arxiv.org/abs/1201.3093}{{\ttfamily arXiv:1201.3093 [nucl-ex]}}.

\bibitem{Adare:2013nff}
{\bfseries PHENIX} Collaboration, A.~Adare {\em et~al.}, ``{Centrality
  categorization for $R_{\rm p(d)A}$ in high-energy collisions},''
  \href{http://dx.doi.org/10.1103/PhysRevC.90.034902}{{\em Phys. Rev.}
  {\bfseries C90} no.~3, (2014) 034902},
\href{http://arxiv.org/abs/1310.4793}{{\ttfamily arXiv:1310.4793 [nucl-ex]}}.

\bibitem{Jamiepriv2}
J.~Nagle {Private communication}, 2017.

\bibitem{Luo:2015doi}
X.~Luo, ``{Exploring the QCD phase structure with Beam Energy Scan in heavy-ion
  collisions},'' \href{http://dx.doi.org/10.1016/j.nuclphysa.2016.03.025}{{\em
  Nucl. Phys.} {\bfseries A956} (2016) 75--82},
\href{http://arxiv.org/abs/1512.09215}{{\ttfamily arXiv:1512.09215 [nucl-ex]}}.

\bibitem{Adamczyk:2017nof}
{\bfseries STAR} Collaboration, L.~Adamczyk {\em et~al.}, ``{Beam energy
  dependence of jet-quenching effects in \AuAu collisions at $\snn$ = 7.7,
  11.5, 14.5, 19.6, 27, 39, and 62.4 GeV},''
\href{http://arxiv.org/abs/1707.01988}{{\ttfamily arXiv:1707.01988 [nucl-ex]}}.

\bibitem{Adam:2016rdg}
{\bfseries ALICE} Collaboration, J.~Adam {\em et~al.}, ``{J/$\psi$ suppression
  at forward rapidity in \PbPb\ collisions at $\snn = 5.02$ TeV},''
  \href{http://dx.doi.org/10.1016/j.physletb.2016.12.064}{{\em Phys. Lett.}
  {\bfseries B766} (2017) 212--224},
\href{http://arxiv.org/abs/1606.08197}{{\ttfamily arXiv:1606.08197 [nucl-ex]}}.

\bibitem{Khachatryan:2016jfl}
{\bfseries CMS} Collaboration, V.~Khachatryan {\em et~al.}, ``{Measurement of
  inclusive jet cross sections in pp and PbPb collisions at $\snn=2.76$ TeV},''
  \href{http://dx.doi.org/10.1103/PhysRevC.96.015202}{{\em Phys. Rev.}
  {\bfseries C96} no.~1, (2017) 015202},
\href{http://arxiv.org/abs/1609.05383}{{\ttfamily arXiv:1609.05383 [nucl-ex]}}.

\bibitem{Dainese:2004te}
A.~Dainese, C.~Loizides, and G.~Paic, ``{Leading-particle suppression in high
  energy nucleus-nucleus collisions},''
  \href{http://dx.doi.org/10.1140/epjc/s2004-02077-x}{{\em Eur. Phys. J.}
  {\bfseries C38} (2005) 461--474},
\href{http://arxiv.org/abs/hep-ph/0406201}{{\ttfamily arXiv:hep-ph/0406201
  [hep-ph]}}.

\bibitem{Zhang:2013oca}
X.~Zhang and J.~Liao, ``{Jet quenching and its azimuthal anisotropy in \AA\ and
  possibly high multiplicity \pA\ and \dA\ collisions},''
\href{http://arxiv.org/abs/1311.5463}{{\ttfamily arXiv:1311.5463 [nucl-th]}}.

\bibitem{Tywoniuk:2014hta}
K.~Tywoniuk, ``{Is there jet quenching in \pPb?},''
\href{http://dx.doi.org/10.1016/j.nuclphysa.2014.04.023}{{\em Nucl. Phys.}
  {\bfseries A926} (2014) 85--91}.

\bibitem{Shen:2016egw}
C.~Shen, C.~Park, J.-F. Paquet, G.~S. Denicol, S.~Jeon, and C.~Gale, ``{Direct
  photon production and jet energy-loss in small systems},''
  \href{http://dx.doi.org/10.1016/j.nuclphysa.2016.02.016}{{\em Nucl. Phys.}
  {\bfseries A956} (2016) 741--744},
\href{http://arxiv.org/abs/1601.03070}{{\ttfamily arXiv:1601.03070 [hep-ph]}}.

\bibitem{Chatrchyan:2013nka}
{\bfseries CMS} Collaboration, S.~Chatrchyan {\em et~al.}, ``{Multiplicity and
  transverse momentum dependence of two- and four-particle correlations in
  \pPb\ and \PbPb\ collisions},''
  \href{http://dx.doi.org/10.1016/j.physletb.2013.06.028}{{\em Phys. Lett.}
  {\bfseries B724} (2013) 213--240},
\href{http://arxiv.org/abs/1305.0609}{{\ttfamily arXiv:1305.0609 [nucl-ex]}}.

\bibitem{Adamczyk:2016gfs}
{\bfseries STAR} Collaboration, L.~Adamczyk {\em et~al.}, ``{Measurement of
  elliptic flow of light nuclei at $\sqrt{s_{NN}}=$ 200, 62.4, 39, 27, 19.6,
  11.5, and 7.7 GeV at the BNL Relativistic Heavy Ion Collider},''
  \href{http://dx.doi.org/10.1103/PhysRevC.94.034908}{{\em Phys. Rev.}
  {\bfseries C94} no.~3, (2016) 034908},
\href{http://arxiv.org/abs/1601.07052}{{\ttfamily arXiv:1601.07052 [nucl-ex]}}.

\bibitem{Loizides:2016tew}
C.~Loizides, ``{Experimental overview on small collision systems at the LHC},''
  \href{http://dx.doi.org/10.1016/j.nuclphysa.2016.04.022}{{\em Nucl. Phys.}
  {\bfseries A956} (2016) 200--207},
\href{http://arxiv.org/abs/1602.09138}{{\ttfamily arXiv:1602.09138 [nucl-ex]}}.

\bibitem{Adam:2015doa}
{\bfseries ALICE} Collaboration, J.~Adam {\em et~al.}, ``{Measurement of jet
  quenching with semi-inclusive hadron-jet distributions in central \PbPb\
  collisions at $\snn=2.76 $ TeV},''
  \href{http://dx.doi.org/10.1007/JHEP09(2015)170}{{\em JHEP} {\bfseries 09}
  (2015) 170},
\href{http://arxiv.org/abs/1506.03984}{{\ttfamily arXiv:1506.03984 [nucl-ex]}}.

\bibitem{Adamczyk:2017yhe}
{\bfseries STAR} Collaboration, L.~Adamczyk {\em et~al.}, ``{Measurements of
  jet quenching with semi-inclusive hadron+jet distributions in \AuAu\
  collisions at $\snn$ = 200 GeV},''
\href{http://arxiv.org/abs/1702.01108}{{\ttfamily arXiv:1702.01108 [nucl-ex]}}.

\end{thebibliography}\endgroup

\newpage
\section*{Appendix}
\begin{figure}[bht!] 
 \centering 
 \includegraphics[width=0.95\linewidth]{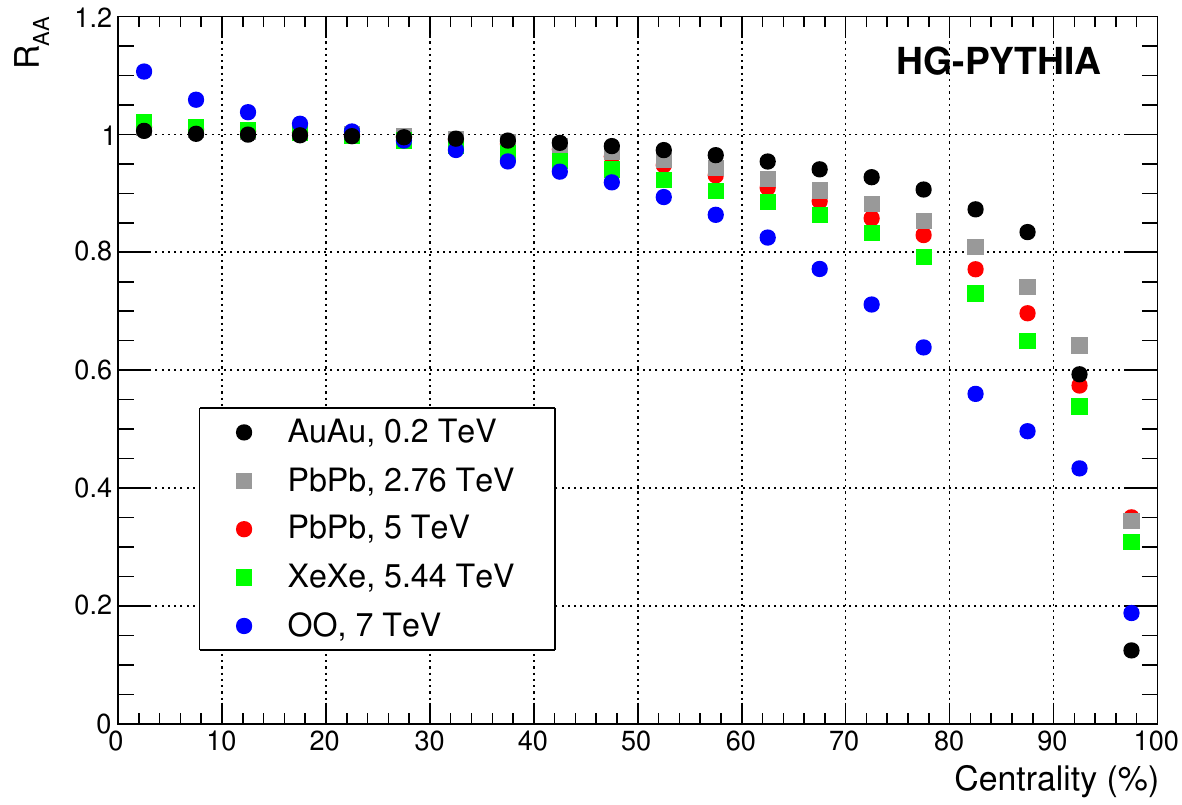}
 \caption{$\RAA$ versus centrality for various collision systems calculated with \Hpythia\ as $\RAA\equiv\Nhard/(\Ncoll\,\left<\Nhardnn\right>)$.}
 \label{fig:app}
\end{figure}

Values for $\RAA$ are provided for numerous collisions systems~(\AuAu\ at 0.2~TeV, \PbPb\ at 2.75 and 5.02~TeV, \XeXe\ at 5.44~TeV and \OO\ at 7~TeV) in \Tab{tab:app} and shown in \Fig{fig:app}.
To obtain the values for other centralities the computed suppression factors must be weighted with $\Ncoll$. If you need values for a different collision system or centrality, please ask the authors or use
the provided code.

\begin{table}[thb!]
  \begin{center}
    \label{tab:values}
    \begin{tabular}{l||r|r|r|r|r}
      \textbf{System} & {\textbf{\AuAu}}  & {\textbf{\PbPb}}  & {\textbf{\PbPb}} & {\textbf{\XeXe}} & {\textbf{\OO}} \\
      $\snn$~(TeV)  & {0.2}  &  {2.76} &  {5.02}  &  {5.44}  &  {7}  \\
      \hline\hline
      Centrality  & $R_{\rm AuAu}$ & $R_{\rm PbPb}$ & $R_{\rm PbPb}$ & $R_{\rm XeXe}$ & $R_{\rm OO}$  \\\hline
    0--5\% & 1.006 & 1.011 & 1.013 & 1.021 & 1.107\\
  5--10\% & 1.001 & 1.007 & 1.008 & 1.013 & 1.059\\
10--15\% & 1.000 & 1.005 & 1.005 & 1.008 & 1.038\\
15--20\% & 0.998 & 1.003 & 1.003 & 1.003 & 1.018\\
20--25\% & 0.997 & 1.000 & 1.000 & 0.997 & 1.005\\
25--30\% & 0.995 & 0.997 & 0.996 & 0.989 & 0.989\\
30--35\% & 0.993 & 0.993 & 0.991 & 0.981 & 0.973\\
35--40\% & 0.989 & 0.987 & 0.984 & 0.969 & 0.954\\
40--45\% & 0.986 & 0.980 & 0.975 & 0.955 & 0.937\\
45--50\% & 0.980 & 0.969 & 0.962 & 0.940 & 0.919\\
50--55\% & 0.973 & 0.957 & 0.948 & 0.923 & 0.894\\
55--60\% & 0.965 & 0.943 & 0.930 & 0.904 & 0.864\\
60--65\% & 0.954 & 0.925 & 0.909 & 0.885 & 0.825\\
65--70\% & 0.941 & 0.905 & 0.887 & 0.864 & 0.772\\
70--75\% & 0.927 & 0.882 & 0.858 & 0.832 & 0.711\\
75--80\% & 0.906 & 0.853 & 0.829 & 0.792 & 0.639\\
80--85\% & 0.873 & 0.809 & 0.771 & 0.730 & 0.560\\
85--90\% & 0.834 & 0.741 & 0.697 & 0.649 & 0.496\\
90--95\% & 0.593 & 0.642 & 0.574 & 0.538 & 0.433\\
95--100\% & 0.124 & 0.345 & 0.350 & 0.309 & 0.188\\\hline

  0--10\% & 1.007 & 1.009 & 1.011 & 1.017 & 1.086\\
10--20\% & 1.002 & 1.004 & 1.004 & 1.006 & 1.029\\
20--30\% & 0.999 & 0.999 & 0.998 & 0.994 & 0.998\\
30--40\% & 0.994 & 0.990 & 0.988 & 0.976 & 0.965\\
40--50\% & 0.986 & 0.975 & 0.971 & 0.949 & 0.929\\
50--60\% & 0.973 & 0.952 & 0.942 & 0.915 & 0.880\\
60--70\% & 0.952 & 0.917 & 0.901 & 0.876 & 0.802\\
70--80\% & 0.922 & 0.871 & 0.847 & 0.816 & 0.677\\
80--90\% & 0.857 & 0.785 & 0.746 & 0.700 & 0.533\\
90--100\% & 0.373 & 0.525 & 0.488 & 0.441 & 0.313\\\hline

  0--20\% & 1.005 & 1.007 & 1.008 & 1.013 & 1.063\\
20--40\% & 0.997 & 0.996 & 0.995 & 0.987 & 0.985\\
40--60\% & 0.982 & 0.968 & 0.961 & 0.938 & 0.909\\
60--80\% & 0.943 & 0.904 & 0.886 & 0.858 & 0.753\\
80--100\% & 0.733 & 0.722 & 0.685 & 0.627 & 0.439\\
    \end{tabular}
  \end{center}
  \caption{Numerical values for $\RAA$ versus centrality for various collision systems calculated with \Hpythia\ as $\RAA\equiv\Nhard/(\Ncoll\,\left<\Nhardnn\right>)$.}
  \label{tab:app}
\end{table}

\end{document}